\documentclass[11pt,journal,onecolumn]{IEEEtran}
\usepackage{graphicx}
\usepackage[american]{babel}
\usepackage{csquotes}
\usepackage{setspace}
\usepackage[backend=biber,style=ieee]{biblatex}
\usepackage{titling}
\usepackage{datetime}

\newdateformat{monthyeardate}{%
  \monthname[\THEMONTH] \THEDAY, \THEYEAR}

\predate{\par\large\centering}
\postdate{\par}

\title{Intuitive control of myoelectric prostheses with agonist-antagonist interface and magnetomicrometery: narrative review
}
\date{}
\author{Ivan R. Slootweg, Radboud University  \\ \monthyeardate\today}

\begin{document}
\maketitle
\section{Abstract}
Proprioception is crucial in intuitive control of  prosthetic limbs  and therefor contributes to intuitive prosthetic use. The agonist-antagonist myoneural interface (AMI) is an prosthetic innovation with enhanced control, reduced pain, and heightened proprioceptive sensation in clinical experiments. Furthermore, studies have  addressed surgical techniques to make this myoelectric interface available to a larger group of patients. A narrative review on AMI's developmental process, surgical implementations, and validation has been conducted for clinical and pre-clinical experimental research, embedded in a theoretical background on feedback control and proprioception. The closing chapter on magnetomicrometery serves to illustrate how motor control reading, as an example component, can benefit from technical innovations to enhance the Intuitive prosthetic control with AMI.
\\
Keywords: agonist-antagonist; myoneural; prosthetic; motor control; proprioception; magnetomicrometery;
\\
\section{Introduction} 
\subsection{Optimal feedback control}
Optimal feedback control describes the mechanism of motor control that combines the computational, biomechanical, sensory and stochastic aspects that affect or are required to control human motor movements \parencite{Sensinger2020}. According to this control model, humans rely on several components for controlling movements: costs and rewards, internal models of mapping between actions and anticipated effects, optimal feedback-driven policy to compute the policy that maximizes the reward, and state estimation that serves as input to the control policy \parencite{Sensinger2020}. This implies that decisions are based on estimated states and not on sensed states. As sensory information has delays and variability caused by noise, it is used to optimize the state estimator retrospectively by making adjustments correlated to the error between predicted and sensed state with Kalman gain. 
\subsection{Somatosensory feedback in prostheses}
Supposedly, the same computational components as in optimal feedback control are involved in the motor control loop of prosthesis users \parencite{Sensinger2020}. However, these components are not as extensively researched as in healthy individuals. The general accepted idea is that the computational methods are similar but prosthesis users have altered internal model uncertainty and therefore different movement control due to altered model parameters such as those involved in the policy and noise \parencite{Johnson2017}. If optimal feedback control is indeed used in prosthesis users, this means that somatosensory feedback is vital in prosthetic limb control. While this hypothesis has been tested in various experiments, results remain inconclusive \parencite{Sensinger2020}.
One explanation for this can be that somatonsory signals that are relied on are inadequately conveyed, augmented or subtituted .
\subsection{Proprioceptive feedback}
Various somatosensory feedback types exist, of which tactile information is mostly facilitated in prostheses, often to substitute other sensory information \parencite{Sensinger2020}. Whilst such implementations have proven to benefit task execution and to strengthen internal models for motor control, performance is not comparable with healthy individuals and still users experience little embodiment and connection with their prosthetic limbs \parencite{Sensinger2020}. As an alternative to tactile information, proprioception informs limb position and movement, forces and balance and consequently is of high value to limb control and paramount for intuitive prosthetic limb control \parencite{Proske2012}. This information is perceived from signal differences between agonists and antagonist muscles. When the agonist muscle contracts, muscle spindle fibers and Golgi tendon organs transduce these forces via the antagonist into afferent signals, allowing for intuitive limb control which in turn is related to a sense of embodiment \parencite{Srinivasan2017, DAnna2019}. Restoring proprioception in prostheses users could potentially improve these aspects, especially because proprioceptive signals contain information on various motor control aspects. The majority of studies on restoring proprioception are on upper limb prostheses, from which various approaches can be derived such as vibrotactile sensory substitution and intraneural stimulation. Unfortunately, any of these methods depends on a mapping from the proprioceptive state to a substitute input signal.
\\ \\
In this review, we will look into the proposed agonist-antagonist myoneural interface (AMI) and the multidisciplinary innovations of this technique, as the first developments of AMI implementations have been rapidly followed by innovative surgical techniques. To leverage the intuitive control of prosthetic limbs that is offered through AMI, faster and more accuracy efferent signal transduction methods could be combined to make one high performance and reliable control interface. Therefore, this review seeks to provide a complete overview of the rationale behind and validation of AMI, including in-human experiments, and the recently proposed technique of magnetomicrometery (MM) in comparison to traditional motor control reading.

\section{Restoring agonist-antagonist dynamics}
Traditional limb amputation surgery creates a padded distal region by layering native muscle tissue around the transected bone, which disrupts the agonist-antagonist dynamic \parencite{Srinivasan2017}. Therefore, to restore the natural proprioceptive sensation in the residual limb, a solution needs to restore this dynamic and should overcome several challenges related to prosthesis control. The first challenge is related to signal transduction of the intended motor command to the prosthesis. As neural recording interfaces are unreliable, solutions like targeted muscle reinnervation (TMR) and regenerative peripheral nerve interface (RPNI) have been proposed to shift the recording objective to the target muscular tissue to amplify the nerve signal. Both methods involve connecting target nerves to muscle grafts \parencite{Srinivasan2017}. Specifically, in TMR transected nerves are connected to surgically denervated muscles in the residuum whereas in RPNI a small muscle graft is reinnervated with a resected peripheral nerve. Although TMR has proven to improve upon traditional techniques, it is too invasive as functional muscles are resected \parencite{Srinivasan2017}. Whilst RPNI solves this problem, it does not allow for agonist-antagonist coupling. Therefore, a technique is sought for that builds upon RPNI with the aim to repair agonist-antagonist dynamics.


\section{AMI}
\subsection{Developmental process}
\subsubsection{The proprioceptive feedback loop}
The idea for maintaining agonist-antagonist dynamics in a residuum after amputation was first proposed in 1896 \parencite{Herr2021}. The procedure, called cineplasty, involved creating skin-lined tunnels of tendon or muscle and connecting these to an prosthetic joint. This allowed the user to directly apply forces to the prosthesis and experience afferent signals from the muscle-tendon stretch in the antagonist muscles. This technique is no longer used because of several disadvantages related to irritation and infection of the tunnels. In 2017, \citeauthor{Clites2017} experimented with a murine model to validate the concept of reading native muscular tissue activity to obtain efferent muscle motor commands and simultaneously delivering sensory information of expected muscle stretch and tension from a prosthesis to afferent nerves \parencite{Clites2017}. Actively delivering this information is needed in case of passive movement by external forces, when agonist activation does not stretch the antagonist muscle, to provide afferent proprioceptive information. Experiments of a positive control were compared between a native agonist-antagonist pair of tibialis anterior and lateral gastrocnemius and the same muscle pair with artificial tendon coupling through an AMI. In these experiments, functional electrical stimulation (FES) stimulated the agonist and electroneurography (ENG) recorded agonist motor afferents with a nerve cuff whilst antagonist stretch was recorded with sonomicrometry crystals. In sonomicrometry, ultrasound time-of-flight of a pair of piezoelectric crystals is used to inform muscle fascicle length measurements. Robustness of these measurements was validated during stretching of the gastroctnemius-soleus complex (GSC) with a muscle tensioner, with sonomicrometry recordings of the GSC and ENG recordings of the afferent tibial nerve. \\
\subsubsection{Surgical implementation techniques}
Shortly after, the same research group proposed a new surgical approach for constructing an AMI through linkage of two muscle grafts and innervating each of such pairs by a pair of transected flexor-extensor motor nerves \parencite{Srinivasan2017}. This experiment aimed to validate the same concept of reading efferent motor commands and delivering afferent proprioceptive information as proposed in aforementioned studies. Motor control signals were read from the agonist by EMG and fascicle state measurements with implanted sensors whereas FES of the antagonist muscle provided feedback on the proprioceptive state of the agonist muscle. Their implementation of AMI was validated in a murine model where transected peroneal and tibial nerves are placed into two separated denervated muscle patches that were linked by their tendons. Following, the first in-human experiments for controlling and sensing proprioception of a prosthesis joint were performed 
\parencite{Clites2018a,Clites2018b}. The very first experiment implemented AMI in three patients and in a second more elaborate evaluation, one subject received two AMIs in their residuum and signals of each AMI were sent to one joint of a two-degree-of-freedom ankle-foot prosthesis \parencite{Clites2018b}. Whilst bipolar surface electrodes adjacent to each involved muscle were used to estimate muscle activation through EMG, afferent proprioception information was provided through FES of the antagonist. This subject was compared in experiments with and without afferent proprioceptive feedback from the prosthesis as well as to four TA's. 
\\ \\ 
After this in-human evaluation of prosthesis control and proprioceptive sensation, future studies aimed at improving on the aspects and addressing limitations of the present techniques. One such limitation is that AMI had been designed only for planned amputations as it required healthy distal tissues, whereas the majority of amputations occur in patients who do not have healthy distal tissue. The main downside of this is related to impaired neural tissue. For one, there is no standardized location for nerve transecting and often nerves are buried deep within the residual tissue. Secondly, AMI requires identification of flexor and extensor nerve pairs, which is often hindered by healing and scarring of nerves. To overcome this, a dual-stage AMI implementation surgery which leverages remaining tissues was proposed \parencite{Srinivasan2019}. The first stage involves dissecting and suturing each terminal nerve in the residual limb into a separate muscle graft. Following this surgery, the patient learns to identify the original function of each graft. In the second stage, tagged grafts inform which agonist-antagonist pairs should be connected into an AMI. Unfavourably, this second stage disrupts the healing process and therefore is expected to affects the functionality of the AMI. Furthermore, it is hypothesised that deploying the tissue of the linked muscle grafts through motor commands affects atrophy in the residuum. To test these assumptions and evaluate the effect of these issues, AMI’s functional components were evaluated in five murine models with the single-stage method, five models with the proposed dual-stage method and a negative control groups in which the second stage was left out. In this murine model, graft tagging was not needed after the first stages as it was known which nerve innervates each graft. Following these innovations towards an AMI within one research group, another team of \citeauthor{Wang2022} addressed a problem that remained unaddressed by \citeauthor{Clites2018a}'s approach of direct agonist-antagonist connection \parencite{Wang2022}. In patients with higher degrees of amputations, there are no or little opportunities for establishing muscle-tendon tissue connections. Furthermore, the researchers argued that the correlation between muscle force and electrical signal is a weak demonstration for the feedback loop in AMI and more details on synergy within the muscle pairs are required. To make AMI available to higher-degree amputees, it was proposed to extend the selection of nerve implantation to the whole body in a method inspired by RPNI, which was demonstrated in a murine model. This involves a new strategy in which a closed structure of nerves, muscle and tendon is created by anastomosing the ends of tendon tissue that is part of a donor muscle and implanting two nerves into the muscle. This approach was recreated in a murine model with a donor soleus muscle and tibial nerve and common peroneal nerve, and compared with simple tendon and nerve dissection. Like other aforementioned studies, efferent myoelectric and afferent neuroelectric signals were used to validate the potential of the feedback loop in the AMI approach. In addition, to demonstrate synergy within the muscle pairs, electrophysiological signaling, animal behavior, and histomorphology were used as biological indicators. 



\subsection{Validation of AMI}
\subsubsection{Fundamental support for proprioceptive feedback loop}
The first murine experiments introduced sonomicrometry as a reliable indicator of muscle length \parencite{Clites2017}. Furthermore, a correlation between muscle stretch and afferent neural signal implicated that muscle stretch is proportional to afferent nerve signals. This supports the proprioceptive feedback loop that is proposed in AMI in which limb position is encoded via afferent nerve paths of the antagonist. The strength of this correlation also confirmed the viability of the tendon coupling technique in AMI muscle pairs. \citeauthor{Srinivasan2017} interpreted the histopathological, electrophysiological and functional properties of AMI \parencite{Srinivasan2017}. They observed successful reinnervation and regeneration of agonist and antagonist muscles and exclusive depolarisation of the target muscle was confirmed. This is important for a good signal to noise ratio. Furthermore, like in \parencite{Clites2017}, agonist and agonist motion was successfully coupled as agonist flexing and antagonist stretching were proportional and EMG signal showed proportional response to agonist stimulation.
\\ \\
\subsubsection{Functional validity of AMI}
The in-human experiments in \parencite{Clites2018a, Clites2018b} allowed for the functional evaluation as well as qualitative evaluation of AMI by users of a myoelectric prosthesis. First, positive surgical outcomes for three patients were described as well as isolated muscle activation, great perception of phantom limb position and no reports of phantom pain \parencite{Clites2018a}. Secondly, in agreement with studies in murine models, agonist activation was correlated with antagonist stretch, both in isolated movements as well as during alternating contractions \parencite{Clites2018b}. This supports the validity of AMI in realistic limb movements. In comparison to TA, subject with AMI exhibited better control over joint placement and improved reflexion during stair walking. Furthermore, tuning of the controller gains allowed the subject to note little delay between intentional motor activation and movement of their prosthesis and they expressed a strong sense of embodiment. These results indicate an AMI brings valuable improvements to prosthetic limb users. In a follow-up experiment, AMI subjects (n=15) reported less pain and could produce higher range of motion compared to traditional amputees (TA) (n=7) subjects \parencite{Srinivasan2021}.
The dual-stage approach introduced in \parencite{Srinivasan2019} had to be evaluated for its effects on muscle functionality and tissue atrophy. Although atrophy occurred between the first and second stage, the grafts had increased in size after stage two. Furthermore, the AMI grafts significantly preserved muscle mass in comparison to unlinked grafts and mechanical and electrophysiological properties confirmed proprioceptive feedback loops. This indicates that a dual-stage approach is a suitable replacement for the traditional AMI technique. Like the other AMI implementations, the proposed method for higher degree amputees significantly improved stability and muscle synergism \parencite{Wang2022}. Furthermore, repair of the motor control loop by neuro feedback was confirmed in AMI by improvements in various gait cycles.  To conclude, these studies all agree on the validity of AMI on a functional motor level. \\
\subsubsection{Restoration of proprioceptive neurophysiology}
After the validation of technical developments, \citeauthor{Srinivasan2020} provided insight into proprioceptive sensorimotor activity in the central nervous system following AMI use \parencite{Srinivasan2020}. Structural and function MRI (fMRI) were acquired in rest and during tasks execution and compared among 12 subjects with AMI amputation, 7 TA and 10 without amputation. TA subjects demonstrated significant lower proprioceptive activity, whilst functional connectivity of sensorimotor and motor regions in the brain of patients with an AMI was similar to controls with no amputation. This demonstrates that AMI can regenerate proprioceptive feedback. Furthermore, the study confirmed the basic working principle of AMI that was evaluated in the earlier fundamental studies on a cognitive level, as the proprioceptive activity in the brain demonstrated strong correlation with muscle activation. Other connectivity analyses in resting state indicated a contrast in information reliance for prosthesis proprioception among the study groups. Whereas TA subjects had the highest coupling of the sensorimotor system by visual areas, users of AMI had lower coupling of these systems an experienced higher phantom sensation. These are considered as indications for preservation of proprioceptive neurophysiology of the lower limb by an AMI. A similar study with fMRI was performed in resting state on the same population \parencite{Chicos2023}. This study had a stronger focus on prosthesis embodiment and limb sensation, which are important factors that can affect pain in amputees and prosthesis users. Lower coupling of salience networks and motor execution networks was observed in AMI subjects compared to TA subjects, indicating a weaker connection between areas responsible for motor control and dictating focus on salient stimuli. This could be related to less reliance of motor control on visual information in AMI users, which was found in \parencite{Srinivasan2020}. In support of this, a weaker connection of the visual cortex and salience networks was observed in AMI subjects. Other support for restored proprioception in AMI users is derived from indications for less information consolidation during movements. More interestingly, other connectivity analyses even indicate significant decrease of neuropathological signatures that are present in TA subjects. These findings of neural plasticity and less reliance on visual input are potential indicators of less cognitive load and improved embodiment in AMI users \parencite{Rackerby2022}. Furthermore, sensorimotor connectivity is argued as an indicator of embodiment as it has been associated with phantom sensation which is related to embodiment \parencite{Giummarra2010}. In support of such findings, \citeauthor{Chicos2023} provide multiple neurophysiological indicators of neuroplasticity in amputees and exploitation of this in AMI subjects, indicated by network reorganisation. 
\section{Reading motor control signals with Magnetomicrometry}
\subsection{Afferent control signals}
The developmental history of AMI has used sonomicrometry and EMG for muscle activity measurements. Such muscle measurements inform the commands for prosthesis control and fast, accurate and high-resolution control is paramount for prosthesis embodiment and reliable proprioception \parencite{Clites2018a,DAnna2019}. For this reason, the development of signal transduction techniques to meet such requirement is of high value for leveraging the proprioceptive information AMI can deliver to prosthesis users and in turn for intuitive prosthetic control. For AMI, such techniques should facilitate observations of efferent nerve intentions in the agonist stream. Whereas wireless surface nerve recordings are not suitable due to the high noise ratio and location deep in the residuum, implantable nerve sensors have biocompatibility problems and bring problems of recording selectivity and noise due to their small size. Therefore, solutions like targeted muscle reinnervation (TMR) and regenerative peripheral nerve interface (RPNI) have been proposed to shift the recording objective to muscular tissue which will amplify the nerve signal. This design approach is reflected in the forward signal loop of AMI, whilst the challenge remains of reliable recordings of the target muscles. Surface EMG (sEMG) has been the most widely adopted approach but is of low quality as the muscles deep in the residuum cannot be targeted consistently through transdermal recordings. 

\subsection{Development of Magnetomicrometry}
The first developments towards MM are found in improvements of magnet tracking algorithms. \citeauthor{8809206} tackled technical issues of MM related to tracking with low time delay and accounting for earth's magnetic field \parencite{8809206}. Magnets with sizes ranging from 2 to 16 mm are reliable measurement objects with errors as low as 1.25$\%$ in 2 mm sized beads. As more magnets and sensors increase the latency and tracking error decreases when more magnets are used simultaneously, there is a trade-off between latency and error. In a follow-up study, the first experiments with MM for wireless tissue length and motion measurements were performed \parencite{Taylor2021}. Real-time muscle length tracking was performed in four turkeys with one implanted pair of 3 mm magnetic beads into the gastrocnemius muscles in the left and right limbs. The two beads seem a good choice in terms of latency trade-off and a size of 3 mm can limit measurement errors \parencite{8809206}. In these experiments, a magnetic field sensory array was mounted to the leg externally for MM measurements. The muscle was mechanically stretched whilst MM and ground truth fluoromicrometry (FM) measurements were compared. Biocompatibility was assessed with histopathological inspection of tissue samples whilst beads migration was assessed with CT recordings. Complementary to findings of previous studies, MM has proven suitable for permanent implantation without adverse tissue reactions and at larger separation distance, beads are resilient to migration. Following the introduction of MM, the same research group addressed the clinical viability of magnetic bead implantation in muscle tissue \parencite{Taylor2022a}. To this end, an improved coating strategy for magnetic bead implants was designed, and a translatable surgical protocol for implantation has been developed. This work focused on evaluating three important aspects: implantation discomfort, beads migration and biocompatibility. The methods were evaluated in running and walking experiments in ten wild turkeys. Comfort was confirmed by absence of stride time changes, minimal beads migration was observed with CT and biocompatibility was confirmed with histopathological examinations in comparison with in vitro and other in situ models. These results indicate that MM can be safely and reliably implanted in humans. In their latest experiment, a fully wireless and portable system was developed to employ MM for in vivo muscle measurements in real use cases \parencite{Taylor2022b}. To this end, MM was employed in less controlled in vivo conditions to assess the robustness of MM during untethered activity with soft tissue artifacts. The methods were evaluated in ten wild turkeys in running experiments and ascent, descent and free roaming movement and MM measurements were compared with FM. Consistent with the aforementioned experiments, MM can track tissue length with submilimeter accuracy and has a strong correlation to ground truth FM measurements of the muscle.


\section{Discussion and Conclusion}
The historical developmental process of AMI is extensively described, however mostly restricted to one specialised lab. As specialised techniques are required for both the surgery and placement of the biophysical controller, it would be good to validate and invest in adequate clinical implementations in other labs. This would create a team of experienced surgeons and technicians that can collaborate on some of the remaining challenges. One challenge is that the biomechanical framework in AMI is limited in reflecting force and torque transmission throughout muscles as occurs in intact limbs. In AMI only one antagonist muscle is connected to the agonist, instead of a multiplicity of muscles and tendons that are connected to one biological joint. Subsequently, proprioceptive feedback is not exactly natural. However, extension of AMI to facilitate multi-muscle dynamics is not always possible as the amount of intact residual nerves is often limited. Nonetheless it would be interesting to compare more naturally AMI concepts with more muscle dynamics with the current AMI implementation. Another issue related to limited natural afferent information is caused by the fact that AMI has not been tested together with tactile information. A combination of somatosensory information is expected to result in even more intuitive prosthetic control. Furthermore, the method for active proprioceptive feedback through FES of the antagonist muscle is sub optimal. Typically, stretch of an antagonist muscle by external forces is not in all cases accompanied with flexion of the agonist muscle. However, the current feedback mechanism in the controller of AMI would require stimulation of the agonist to create stretch of the agonist in cases the position of the prosthetic limb would require this. This results in the user experiencing muscle contraction as the agonist and antagonist muscles are innervated. Therefore, it is interesting to invest in other methods that can deliver a sensation of antagonist tendon stretch and torque. 
\\ \\
Whereas several cineplasty-like  architectures have been developed as a control bridge between a prosthesis and its user, such techniques require a lot of intact residual tissue and relatively bulky hardware systems \parencite{Kontogiannopoulos_2020}.
MM can deliver wireless muscle measurements with high resolution when compared to standard techniques such as EMG. However, it requires compensating for local deviations of the magnetic field. Sonomicrometry has existed for a much longer time than MM and has been used for various biomedical research purposes. Based on its extensive use, sonomicrometry seems a very suitable technique for in-human measurements for prosthesis control. However, indications of adverse affects, although inconclusive, from the implanted crystals bring danger to using it for permanent implantation in humans \parencite{Korinek_2007}. As MM has been developed to be safely implanted in human, it is the best available technique at this moment. 
\\ \\
Efforts towards intuitive proprioceptive feedback with AMI and accurate control through MM are needed to reduce internal model uncertainty. As a direct consequence, prosthesis users will have to rely less on other input signals, such as visual information. Another consequence of improved model uncertainty is to create better estimates in the optimal feedback control loop, which improves motor command outputs. This in turn will have optimal feedback control parameters more closely resemble that of healthy individuals and users will experience their limb with better embodiment.
\\ \\
To conclude, AMI and MM have been widely researched but not yet in one prosthetic limb. While these developments have progressed independently but virtually in parallel, MM seems to come at the right time as AMI has set the next step for innovative and intuitive control of prosthetic limbs. With MM, control signals can be sent even faster and safer to users.
\newpage
\printbibliography

\end{document}